\documentclass[a4paper,twocolumn]{esapub2005} 
\usepackage{times}
\usepackage{natbib}
\usepackage{graphicx}
\newcommand{\beq}{\begin{equation}}
\newcommand{\eeq}{\end{equation}}
\newcommand{\beqnn}{\begin{displaymath}}	
\newcommand{\eeqnn}{\end{displaymath}}		
\newcommand{\beqa}{\begin{eqnarray}}
\newcommand{\eeqa}{\end{eqnarray}}
\newcommand{\beqann}{\begin{eqnarray*}}
\newcommand{\eeqann}{\end{eqnarray*}}

\newcommand{\ben}{\begin{enumerate}}
\newcommand{\een}{\end{enumerate}}
\newcommand{\bit}{\begin{itemize}}
\newcommand{\eit}{\end{itemize}}
\newcommand{\bc}{\begin{center}}
\newcommand{\ec}{\end{center}}

\newcommand{\eqref}[1]{Equation~\ref{#1}}

\newcommand{\eboo}{\mbox{$\eta$~Boo}}

\def\Msol{\mbox{${M}_\odot$}}

\def\la{\mathrel{\hbox{\rlap{\hbox{\lower4pt\hbox{$\sim$}}}\hbox{$<$}}}}
\def\ga{\mathrel{\hbox{\rlap{\hbox{\lower4pt\hbox{$\sim$}}}\hbox{$>$}}}}

\def\laeq{\lower.5ex\hbox{{$\:\scriptstyle\buildrel < \over \sim\:$}}}
\def\gaeq{\lower.5ex\hbox{{$\:\scriptstyle\buildrel > \over \sim\:$}}}

\makeatletter 
 \def\sub#1{\relax\ifmmode _{\fam\z@ #1}\else
         $_{\fam\z@ #1}$\fi}
 \def\super#1{\relax\ifmmode ^{\fam\z@ #1}\else
         $^{\fam\z@ #1}$\fi}
\makeatother

\newcommand{\mycaption}[3]
{\if*#2 \caption{#3\label{#1}}
 \else  \caption[#2]{#3\label{#1}}
 \fi}

\newcommand{\comment}[1]{\relax}

\long\def\COMMENT#1\ENDCOMMENT{}
\def\ENDCOMMENT{}

\newcommand{\Dnu}[1]{\Delta \nu_{#1}}

\newcommand{\muHz}{\mbox{$\mu$Hz}}

\newcommand{\acena}{\mbox{$\alpha$~Cen~A}}
\newcommand{\acenb}{\mbox{$\alpha$~Cen~B}}
\newcommand{\acen}{\mbox{$\alpha$~Cen}}

\newcommand{\bhyi}{\mbox{$\beta$~Hyi}}
\newcommand{\nuind}{\mbox{$\nu$~Ind}}

\newcommand{\bvir}{\mbox{$\beta$~Vir}}
\newcommand{\muara}{\mbox{$\mu$~Ara}}

\title{Observations of solar-like oscillations}
\author{Timothy R. Bedding}
\affil{School of Physics A28, University of Sydney, NSW 2006, Australia}
\author{Hans Kjeldsen}
\affil{Department of Physics and Astronomy, University of Aarhus,
DK-8000 Aarhus C, Denmark}

\begin{document}


\maketitle

\begin{abstract}
There has been tremendous progress in observing oscillations in solar-type
stars.  In a few short years we have moved from ambiguous detections to
firm measurements.  We review the recent results, most of which have come
from high-precision Doppler measurements.  We also review briefly the
results on giant and supergiant stars and the prospects for the future.
\end{abstract}

\section{Main-sequence and subgiant stars}

\begin{figure}
\centering
\includegraphics[width=\linewidth]{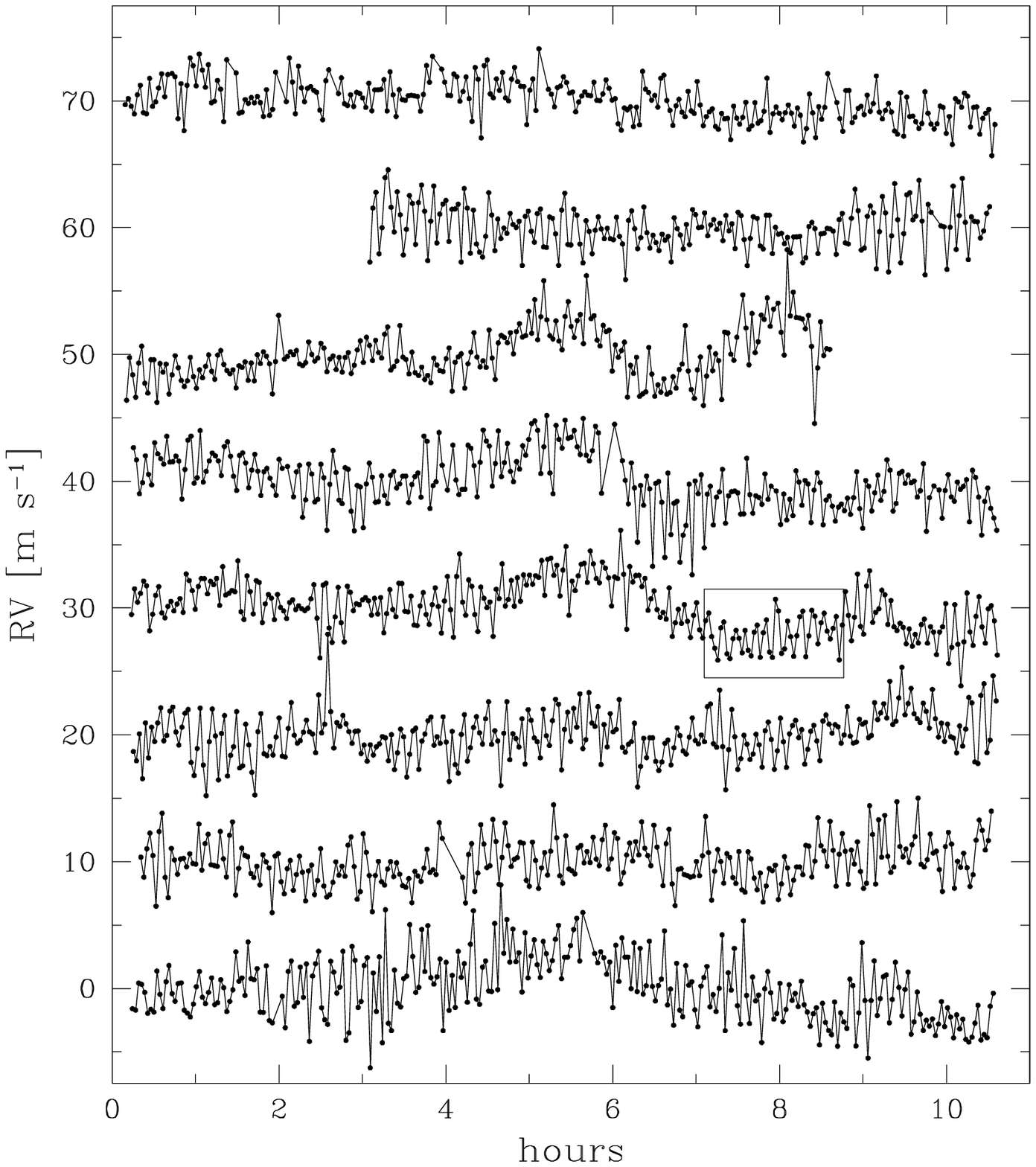}
\caption{Radial velocity time series of the star \muara{} made over 8
  nights with the HARPS spectrograph.  Figure from
  \citet{BBS2005}.\label{fig.muara}}


\bigskip

\bigskip

\includegraphics[width=\linewidth]{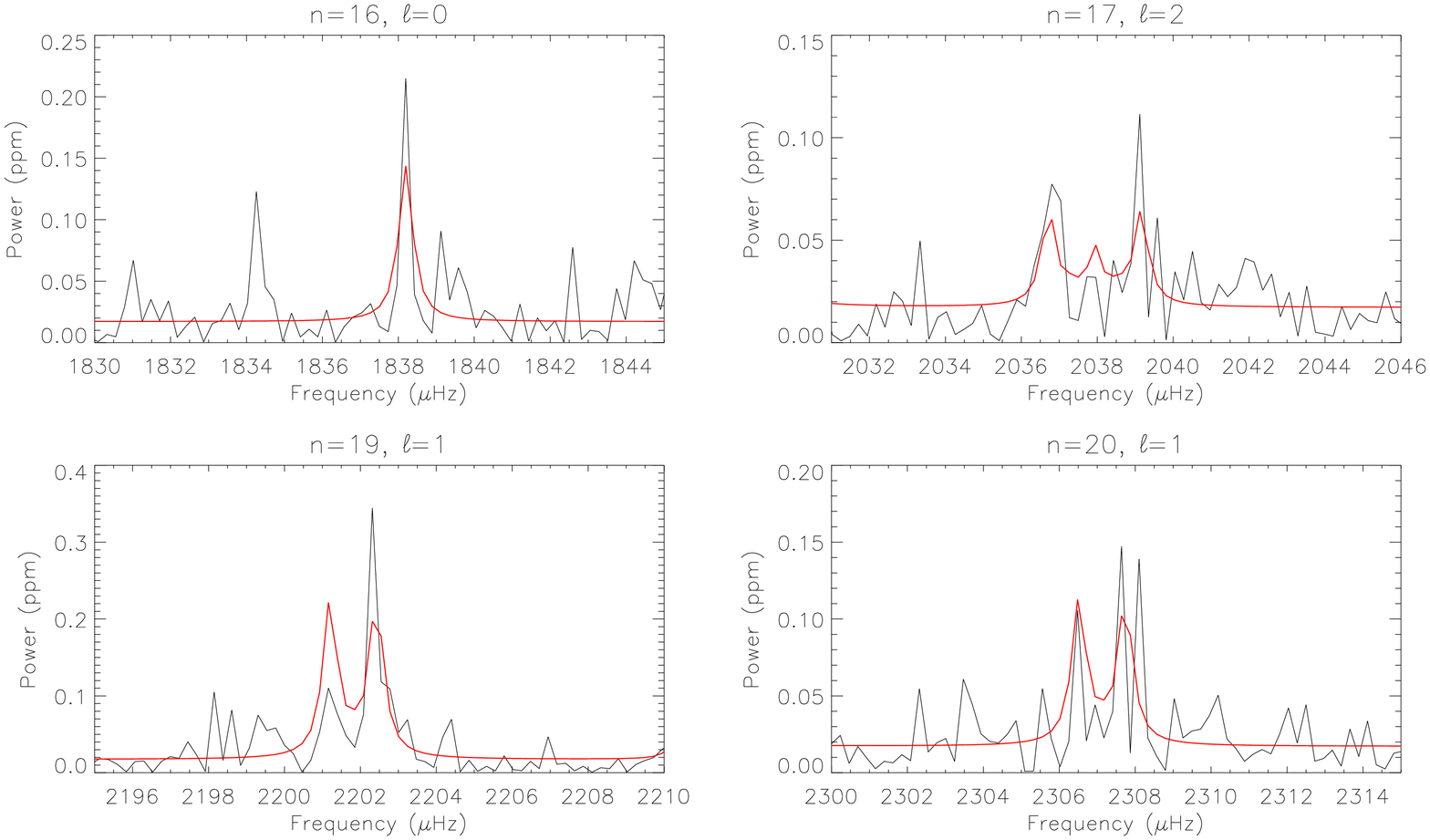}
\caption{Four oscillation modes in \acena{} from the WIRE power spectrum,
  with fits that indicate the linewidth and rotational splitting.  Figure
  from \citet{FCE2006}.\label{fig.fletcher}}


\bigskip

\bigskip

\includegraphics[width=\linewidth]{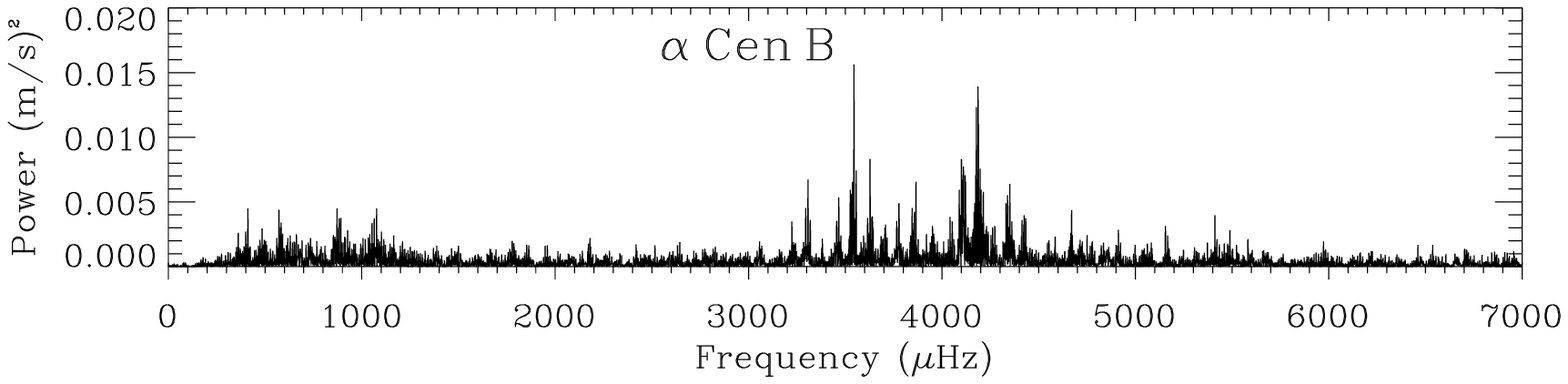}
\caption{Power spectrum of \acenb{} from velocity observations.  Note the
  double-humped structure with a central dip.  Figure from
  \citet{KBB2005}.\label{fig.acenb}}
\end{figure}

There has been tremendous progress in observing oscillations in solar-type
stars, lying on or just above the Main Sequence.  In a few short years we
have moved from ambiguous detections to firm measurements.  Most of the
recent results have come from high-precision Doppler measurements using
spectrographs such as CORALIE, HARPS, UCLES and UVES (see
Fig.~\ref{fig.muara} for an example).  The best data have be obtained from
two-site campaigns, although single-site observations are also being
carried out.  Meanwhile, photometry from space gives a much better
observing window than is usually achieved from the ground but the
signal-to-noise is poorer.  The WIRE and MOST missions have reported
oscillations in several stars, although not without controversy, as
discussed below.

The following list includes all recent observations of main-sequence and
subgiant stars of which we are aware.  The list is ordered according to
decreasing stellar density (i.e., decreasing large frequency
separation,~$\Dnu{}$):
\begin{itemize}  \itemsep=0pt

\item $\tau$~Cet (G8 V): this star was observed with HARPS by
T. C. Teixeira et al.\ (in prep.)

\item 70 Oph A (K0 V): this is the main component of a spectroscopic visual
binary (the other component is K5~V).  It was observed over 6 nights with
HARPS by \citet{C+E2006}, who found $\Dnu{} = 162$\,\muHz{} but were not
able to give unambiguous mode identifications from these single-site data.

\item \acena{} and~B (G2~V and K1~V): see Sec.~\ref{sec.acen}.

\item \muara{} (G3 V): this star has multiple planets.  Oscillations were
  measured over 8 nights using HARPS by \citet{BBS2005} (see
  Fig.~\ref{fig.muara}) and the results were modelled by \citet{BVB2005}.
  They found $\Dnu{} = 90$\,\muHz{} and identified over 40 frequencies,
  with possible evidence for rotational splitting.

\item HD~49933 (F5~V): this is a potential target for the COROT space
mission and was observed over 10 nights with HARPS by \citet{MBC2005}.
They reported a surprisingly high level of velocity variability on
timescales of a few days.  This was also present as line-profile variations
and is therefore presumably due to stellar activity.  The observations
showed excess power from p-mode oscillations and the authors determined the
large separation ($\Dnu{} = 89\,\muHz$) but were not able to extract
individual frequencies.

\item \bvir{} (F9 V): oscillations in this star were detected in a
  weather-affected two-site campaign with ELODIE and FEROS by
  \citet{MLA2004b}.  Subsequently, \citet{CEDAl2005} used CORALIE with good
  weather but a single site, and reported 31 individual frequencies.  Those
  results were modelled by \citet{E+C2006}, who also reported tentative
  evidence for rotational splittings.  The large separation is 72\,\muHz.

\item Procyon A (F5 IV): see Sec.~\ref{sec.procyon}.

\item \bhyi{} (G2 IV): oscillations were detected in \bhyi{} in 2001 using
  UCLES \citep{BBK2001} and CORALIE \citep{CBK2001}.  This star was the
  target for a two-site campaign in 2005, with HARPS and UCLES, that
  resulted in the clear detection of mixed modes (Bedding et al., in
  prep.).  The large separation is 57.5\,\muHz.

\item $\delta$~Eri (K0 IV): \citet{CBE2003a} observed this star over 12
  nights in 2001 with CORALIE and found a large separation of 44\,\muHz.

\item \eboo\ (G0 IV): see Sec.~\ref{sec.eboo}.

\item \nuind\ (G0 IV): this a metal-poor subgiant ($\mbox{[Fe/H]} = -1.4$)
  which was observed from two sites using UCLES and CORALIE\@. The large
  separation of 24\,\muHz, combined with the position of the star in the
  H-R diagram, indicated that the star has a low mass a low mass
  ($0.85\pm0.04\,\Msol$) and is at least 9\,Gyr old \cite{BBC2006}.
  Further analysis is underway (Carrier et al., in prep.).

\end{itemize}

\subsection{\acena{} and B} \label{sec.acen}

On the main-sequence, the most spectacular results have been obtained for
the \acen{} system.  The clear detection of p-mode oscillations in \acena{}
by \citet{B+C2002} using the CORALIE spectrograph represented a key moment
in this field.  This was followed by a dual-site campaign on this star with
UVES and UCLES \citep{BBK2004} that yielded more than 40 modes, with
angular degrees of $l=0$ to~$3$ \citep{BKB2004}.  The mode lifetime is
about 2--4 days and there is now evidence of rotational splitting from
photometry with the WIRE satellite analysed by \citet{FCE2006} (see
Fig.~\ref{fig.fletcher}) and also from ground-based spectroscopy with HARPS
\citep{BBK2006}.

Meanwhile, oscillations in the B component were detected from single-site
observations with CORALIE by \citet{C+B2003}.  Dual-site observations with
UVES and UCLES (see Fig.~\ref{fig.acenb}) allowed measurement of nearly 40
modes and of the mode lifetime \citep{KBB2005}.

We have previously pointed out \citep{B+K2006} that the power spectrum of
Procyon appears to show a dip at 1.0\,mHz that is apparently consistent
with the theoretical models of \citet{HBChD99}.  A similar dip for low-mass
stars was also discussed by Houdek at this conference (see also
\citet{CEH2006}), and the observations of \acenb{} in Fig.~\ref{fig.acenb}
do indeed show such a dip, although not at the frequency indicated by the
models.  It seems that the shape of the oscillation envelope is an
interesting observable that can be extracted from the power spectrum and
compared with theoretical models.

\subsection{\eboo}  \label{sec.eboo}

This star, being the brightest G-type subgiant in the sky, remains a very
interesting target.  The claimed detection of oscillations almost decade
ago by \citet{KBV95}, based on fluctuations in Balmer-line
equivalent-widths, has now been confirmed by further equivalent-width and
velocity measurements by the same group \citep{KBB2003} and also by
independent velocity measurements with the CORALIE spectrograph
\citep{CEB2005}.  With the benefit of hindsight, we can now say that
\eboo{} was the first star for which the large separation and individual
frequencies were measured.  However, there is still disagreement on
some of the individual frequencies, which reflects the subjective way in
which genuine oscillation modes must be chosen from noise peaks and
corrected for daily aliases.  Fortunately, the large separation is $\Dnu=
40$\,\muHz, which is half way between integral multiples of the
11.57-\muHz{} daily splitting (40/11.57 = 3.5).  Even so, daily aliases are
problematic, especially because some of the modes in \eboo{} appear to be
shifted by avoided crossings.

The first spaced-based observations of \eboo, made with the MOST satellite,
have generated considerable controversy.  \citet{GKR2005} showed an
amplitude spectrum (their Fig.~1) that rises towards low frequencies in a
fashion that is typical of noise from instrumental and stellar sources.
However, they assessed the significance of individual peaks by their
strength relative to a fixed horizontal threshold, which naturally led them
to assign high significance to peaks at low frequency.  They did find a few
peaks around 600\,\muHz{} that agreed with the ground-based data, but they
also identified eight of the many peaks at much lower frequency
(130--500\,\muHz), in the region of rising power, as being due to
low-overtone p-modes.  Those peaks do line up quite well with the regular
40\,\muHz{} spacing, but extreme caution is needed before these peaks are
accepted as genuine.  This is especially true given that the orbital
frequency of the spacecraft (164.3\,\muHz) is, by bad luck, close to four
times the large separation of \eboo{} (164.3/40 = 4.1).  Models of \eboo{}
based on the combination of MOST and ground-based frequencies have been
made by \citet{SDG2006}.

\subsection{Procyon}  \label{sec.procyon}

Procyon has long been a favourite target for oscillation searches.  There
have been at least eight separate velocity studies, mostly single-site,
that have reported a hump of excess power around 0.5--1.5\,mHz.  See
\citet{MLA2004}, \citet{BMM2004} and \citet{CBL2005} for the most recent
examples.  However, there is not yet agreement on the oscillation
frequencies, although a consensus is emerging that the large separation is
about 55\,\muHz.  

This star generated controversy when MOST data reported by \citet{MKG2004}
failed to reveal oscillations that were claimed from ground-based data.
However, \citet{BKB2005} argued that the MOST non-detection was consistent
with the ground-based data.  Using space-based photometry with the WIRE
satellite, \citet{BKB2005b} extracted parameters for the stellar granulation
and found evidence for an excess due to p-mode oscillations.

A multi-site campaign on Procyon is being organised for January 2007, which
will be the most extensive velocity campaign so far organised on a
solar-type oscillator.

\section{G and K giants}

There have been detections of oscillations in red giant stars with
oscillation periods of 2--4 hours.  Ground-based velocity observations were
presented at the last SOHO/GONG meeting by \citet{BdRM2004}, who used
CORALIE and ELODIE spectrographs to find excess power and a possible large
separation for both $\epsilon$~Oph (G9 III) and $\eta$~Ser (K0 III).  The
data for $\epsilon$~Oph have now been published by \citet{DeRBC2006}.
\citet{HADeR2006} have analysed the line-profile variations and found
evidence for non-radial oscillations.

Meanwhile, earlier observations of oscillations in $\xi$~Hya (G7 III) by
\citet{FCA2002} have been further analysed by \citet{SKB2004}, who found
evidence that the mode lifetime is only about 2\,days.  If confirmed, this
would significantly limit the the prospects for asteroseismology on red
giants.

\section{Red giants and supergiants}

If we define solar-like oscillations to be those excited and damped by
convection then we expect to see such oscillations in all stars on the cool
side of the instability strip.  Evidence for solar-like oscillations in
semiregular variables, based on visual observations by groups such as the
AAVSO, has already been reported. This was based on the amplitude
variability of these stars \citep{ChDKM2001} and on the Lorenztian profiles
of the power spectra \citep{Bed2003,BKK2005}.

\begin{figure}
\centering
\includegraphics[width=0.9\linewidth]{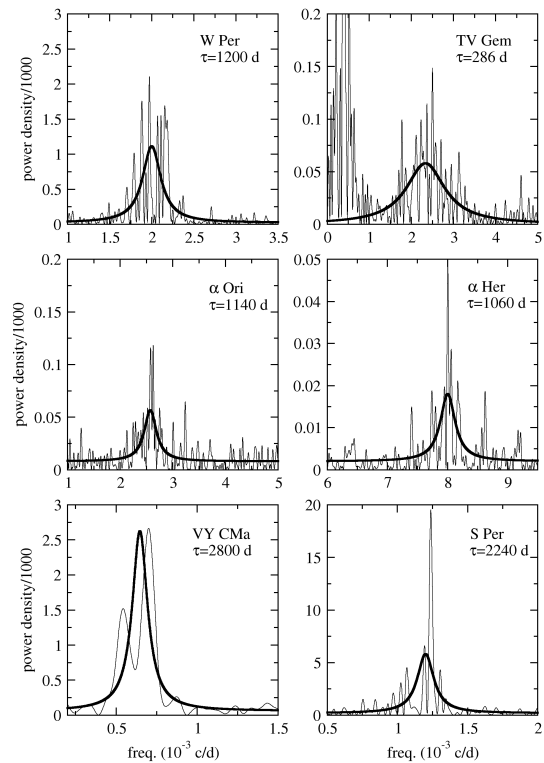}
\caption{Power spectra of red supergiants from visual observations (thin
  lines) with Lorentzian fits (thick lines).  Figure from
  \citet{KSB2006}.\label{fig.kiss}}
\end{figure}

Recently, \citet{KSB2006} used visual observations from the AAVSO to show
that red supergiants, which have masses of 10--30\Msol, also have
Lorenztian profiles in their power spectra (see Fig.~\ref{fig.kiss}).

\section{The future}

In the future, we expect further ground-based observations using Doppler
techniques (for example, a multi-site campaign on Procyon has been
organized for January 2007).  The new spectrograph SOPHIE at l'Observatoire
de Haute-Provence in France should be operating very soon ({\tt\small
http://www.obs-hp.fr/}).  From space, the WIRE and MOST satellites continue
to return data and we look forward with excitement to the expected launches
of COROT (November 2006) and Kepler (2008).

Looking further ahead, the SIAMOIS spectrograph is planned for Dome C in
Antarctica (Seismic Interferometer Aiming to Measure Oscillations in the
Interior of Stars; see {\tt\small http://siamois.obspm.fr/}).  Finally,
there are ambitious plans to build SONG (Stellar Oscillations Network
Group), which will be a global network of small telescopes equipped with
high-resolution spectrographs and dedicated to asteroseismology and planet
searches (see {\small\tt http://astro.phys.au.dk/SONG}).

\section*{Acknowledgments}

This work was supported financially by the Australian Research Council, the
Science Foundation for Physics at the University of Sydney, the Danish
Natural Science Research Council, and the Danish National Research
Foundation through its establishment of the Theoretical Astrophysics
Center.


\end{document}